\documentclass[prd,aps,twocolumn,floats,floatfix,nofootinbib] {revtex4-1}

\usepackage{graphicx}
\usepackage{dcolumn}
\usepackage{bm}
\usepackage{graphics}
\usepackage{slashed}
\usepackage{amssymb}
\usepackage{natbib}
\usepackage{amsmath}
\usepackage{url}
\usepackage[usenames,dvipsnames,svgnames,table]{xcolor}
\usepackage{color}
\usepackage{xcolor}
\usepackage{tabularx}
\usepackage{hyperref}

\begin{document}

\title{Gravitational wave signatures of dark matter cores in binary neutron star mergers by using numerical simulations}

\author{
Miguel Bezares$^{1}$,
Daniele Vigan\`o$^{1}$,
Carlos Palenzuela$^{1}$
}

\affiliation{${^1}$Departament  de  F\'{\i}sica $\&$ IAC3,  Universitat  de  les  Illes  Balears  and  Institut  d'Estudis
Espacials  de  Catalunya,  Palma  de  Mallorca,  Baleares  E-07122,  Spain}

\begin{abstract}

Recent detections by the gravitational wave facilities LIGO/Virgo have opened a window to study the internal structure of neutron stars through the gravitational waves emitted during their coalescence. In this work we explore, through numerical simulations, the gravitational radiation produced by the merger of binary neutron stars with dark matter particles trapped on their interior, focusing on distinguishable imprints produced by these dark matter cores. Our results reveal the presence of a strong $m=1$ mode in the waveforms during the post-merger stage, together with other relevant features. Comparison of our results with observations might allow us to constraint the amount of dark matter in the interior of neutron star.
\end{abstract}

\maketitle

\section{Introduction}

The detections of gravitational waves (GWs) in the last years by the LIGO and Virgo interferometric observatories, consistent with the merger of binary black hole systems~\cite{GW150914,GW151226,GW170104,GW170608,GW170814,2018arXiv181112907T}, has opened a new era of GW astronomy leading to unprecedented discoveries. More recently, GWs from the inspiral of a binary neutron star (NS) system (GW170817) have been observed by LIGO/Virgo~\cite{PhysRevLett.119.161101,Abbott_2017}, followed by several electromagnetic (EM) counterparts: a gamma-ray burst, GRB170817A~\cite{Abbott_2017_2}, and a thermal infrared/optical spectra consistent with a kilonova~\cite{Abbott_2017_3}. These simultaneous EM and GW observations started a fruitful era of multi-messenger astronomy, which will inevitably lead to breakthroughs in our understanding of some of the most exciting objects and phenomena in the universe. The updated detectors are expected to unveil more of binary NS mergers, with a couple of candidates already detected in GW only during the first month of the O3 operations (see the real-time updated GraceDB database~\cite{GraceDB}).

On a much larger scale, there is overwhelming evidence of the existence of dark matter (DM) in the Universe: the mismatch between the luminosity-inferred matter and the rotational curves in galaxies, the inferred mass distributions in galaxy clusters and in galaxy mergers, and the precise measurements of the cosmological baryonic fraction~\cite{Sumner2002,BERTONE2005279}. Measurements of the matter density and its baryonic component imply that the DM represents about $25\%$ of the total content in the Universe~\cite{doi:10.1146/annurev-astro-082708-101659,refId0}. Many theories have been proposed to account for these non-emitting matter, but the most accepted ones describe DM as particles weakly interacting, with mass ranging from $100$ GeV to several TeV~\cite{Roszkowski_2018}. 

Despite the poor knowledge of the DM-baryon interaction (see for instance the experimental upper limits constraints for  weakly interacting massive particles in~\cite{PhysRevLett.107.131302}), these DM particles have been proposed to cluster relatively more easily in dense stars. In this scenario, due to its orbital motion, a star will sweep through the Galactic DM halo and eventually capture some of the particles on its way~\cite{G_ver_2014}. Despite the surface area of a typical NS being much smaller than other stars, two properties make it very efficient in capturing galactic DM particles~\cite{PhysRevLett.108.191301}. First, the high baryonic density inside a NS provides a much higher probability for DM particles to interact and lose energy, compared to other stars. As a matter of fact, for a given star, the particle will interact if the cross-section of the DM-baryon interaction, $\sigma_{\rm DM},$ is at least of the order of the typical projected area occupied by each baryon, $m_p\,R^{2}_{\star}/M_{\star} \sim m_p /\rho R_{\star}$, which for a NS means $\sigma_{\rm DM}\gtrsim 6 \times 10^{-46}$ cm$^2$ (while for the Sun is ten orders of magnitude larger). Second, the strong gravitational force prevents most DM particles escaping from a NS once it loses some of its energy through interactions. Given enough time (of the order and probably much larger than $10^5$ years~\cite{G_ver_2014}), a NS can capture enough number of DM particles to affect its observational properties, which may then be used to constrain the nature of DM. 

On one hand, if DM particles are self-annihilating, this process modifies the thermal evolution of the NS and could be observed as a bright EM emission of old NS, since the released energy due to the annihilation inside the NS can increase the temperature beyond its natural value~\cite{PhysRevD.77.023006}. On the other hand, if DM particles do not self-annihilate, they might cluster in a small region at the center of the NS, increasing its compactness and changing its internal structure~\cite{CIARCELLUTI201119}. Ultimately, this clustering could even lead to a gravitational collapse~\cite{PhysRevD.40.3221}. Either way, NSs might be therefore sensitive to indirect probes of the presence of DM, and can be used to set constraints both on the density and on the physical properties of DM.

Recently, it was suggested the idea that DM might leave a distinct signature on the GW signal radiated during the coalescence of binary NSs, especially during the post-merger stage~\cite{ELLIS2018607}. The collision between a NS and a star made of axions, one of the most popular DM type candidates~\cite{MARSH20161,10.1093/mnras/stv1050}, was also studied~\cite{10.1093/mnras/sty3158}, showing that future observations might be able to detect such mergers and the signals could enhance our understanding of DM. It has also been suggested that DM may ignite supernovae by the formation and self-gravitational collapse of a DM core~\cite{2019arXiv190500395J}. 
 
Here, we present the first fully relativistic 3D numerical simulations of the merger of binary NSs with DM particles trapped on their interior. We describe these systems by modeling the fermionic component with a perfect fluid and the bosonic DM with a complex scalar field~\cite{suspalen,brito}. The resulting objects, known as fermion-boson stars (FBSs)~\cite{Henriques2005}, allow only a coupling between the boson and the fermion particles through gravity. Notice that these kind of systems have been modeled in different ways in the past. Possible changes in the structure of the star by the presence of (non self-annihilating) DM have also been investigated using a two-fluid model~\cite{PhysRevD.84.107301}. However, we find it more convenient to describe these systems by using FBSs, since the bosonic DM particles might form as a Bose-Einstein condensate which can be represented with a single complex scalar field. Notice also that current observations already set some bounds on the amount of DM particles inside NSs for different DM models~\cite{kou}. The effect of weakly-interacting DM on the structure of the star will be stronger in non-linear dynamical scenarios like the coalescence of two NSs. As we will see later, our simulations reveal that the presence of DM cores leaves a distinct imprint in the GWs during the post-merger phase.
Notice however that,  as it was pointed out in~\cite{ELLIS2018607}, in a standard scenario only a small amount of	weakly interacting massive particles DM might cluster in the interior of NSs, accounting for a tiny fraction of the total mass of the star. In this work we will consider heavier DM cores, with up to $10\%$ of the total mass, in order to set upper bounds on the possible effects of these cores on the dynamics. 

This work is organized as follows. In Section~\ref{setup} a brief introduction of the evolution equations describing FBSs is presented, followed by the evolution formalism and numerical implementation. The construction of initial data for FBSs, either isolated or in binaries, is described extensively in Section~\ref{id_fb}. In Section~\ref{coales_fb}, we study the dynamics and the gravitational radiation produced during the binary FBSs coalescence. Finally, we discuss our results in Section~\ref{discu_fb}. We have chosen geometric units such that $G=c=M_{\odot}=1$ and we adopt the convention where roman indices $a,b,c,\ldots$ denote spacetime components (i.e., from $0$ to $3$), while $i,j,k,\dots$ denote spatial ones.
\section{Setup}\label{setup}

Binary NSs with a fraction of DM on their interiors can be modeled by using two different matter components: a perfect fluid for the fermionic matter, and a complex scalar field for effectively describing the bosonic DM. Stationary compact solutions of such  systems are known in the literature as FBSs~\cite{suspalen,Henriques2005}.  
In this section, we present the Einstein-Klein-Gordon-Hydrodynamics (EKGH) system of equations, describing the coupled evolution of the spacetime and the fermionic and bosonic matter components. We also briefly describe the numerical implementation and the quantities used for analyzing the dynamics.

\subsection{Evolution equations}
The spacetime geometry is described by the Einstein equations, which can be extended in a convenient way by using the covariant conformal Z4 formulation (CCZ4)~\cite{alic,bezpalen}, namely
\begin{eqnarray}
   R_{ab} + \nabla_a Z_b + && \nabla_b Z_a = 
   8\pi \, \left( T_{ab} - \frac{1}{2}g_{ab} \,{\rm tr} T \right)\nonumber \\
  && + \kappa_{z} \, \left(  n_a Z_b + n_b Z_a - g_{ab} n^c Z_c \right)\,,
\label{Z4cov}
\end{eqnarray}
where $R_{ab}$ is the Ricci tensor associated to the spacetime metric $g_{ab}$ and $T_{ab}$ is the total stress-energy tensor with trace ${\rm tr} T \equiv g^{ab} T_{ab}$. For this particular problem it can be decomposed as
\begin{eqnarray}
T_{ab}= T^{\Phi}_{ab} + T^{M}_{ab}~,
\label{FB_tensor}
\end{eqnarray}
where $T^{\Phi}_{ab}$ is the energy-momentum tensor associated to a (complex) scalar field and $T^{M}_{ab}$ corresponds to a perfect fluid. The $Z_{a}$ four-vector measures deviations from Einstein's solutions~\cite{Z41,Z42} (i.e., those satisfying the constraints). Notice that suitable damping terms, proportional to the parameter $\kappa_{z}>0$, have been included in order to enforce the dynamical decay of the constraint violations associated with $Z_{a}$~\cite{gundlach,bezpalen}.

\subsubsection{Bosonic matter}

We consider here stable stars (i.e., not collapsing to black holes) such that only a relatively small amount of DM dwells in the core of each NS. Each DM core is described by a complex scalar field, $\Phi^{(i)}$, and its energy-momentum tensor is
\begin{eqnarray}
T^{(i)}_{ab} &=& \nabla_a \Phi^{(i)} \nabla_b \bar{\Phi}^{(i)} +    \nabla_a \bar{\Phi}^{(i)} \nabla_b \Phi^{(i)} \nonumber   \\
&& ~ -g_{ab} \left[ \nabla^c \Phi^{(i)} \nabla_c \bar{\Phi}^{(i)}   + V^{(i)}\left(\left|\Phi^{(i)}\right|^2\right) \right]~,
	\label{DS_tensor}  
\end{eqnarray}  
being $\bar{\Phi}^{(i)}$ the complex conjugate of $\Phi^{(i)},$  $V^{(i)}(|\Phi^{(i)}|^2)$ the scalar field potential and the super-script $(i)$ denotes each star. The total energy-momentum tensor for the bosonic components is just the superposition of both
stars, that is,
\begin{equation}
T^{\Phi}_{ab} = T^{(1)}_{ab} + T^{(2)}_{ab}~.
\end{equation}
The scalar fields evolve according to the Klein-Gordon (KG) equations
\begin{eqnarray}
g^{ab} \nabla_a \nabla_b \Phi^{(i)} &=& \frac{dV^{(i)}}{d \left|\Phi^{(i)}\right|^2} \Phi^{(i)}~.\label{FB_EKG1}
\end{eqnarray}
In this work we will consider the simplest self-potential for each scalar field, given by 
\begin{equation}
V\left(|\Phi^{(i)}|^2\right)=\mu^2|\Phi^{(i)}|^2\,,\label{potential}
\end{equation}
where $\mu$ is a free parameter related to boson mass and the $|\Phi|^{2}$ dependence ensures the U(1) invariance of the Lagrangian. This interaction potential, for isolated DM cores, leads to the well-known \textit{mini-Boson star} (see \cite{liebpa} for an overview of different kind of potentials). One can use the conserved current
\begin{equation}
J^{(i)}_{a} = i g_{ab} \left(\bar{\Phi}^{(i)}\,\nabla^{b}\Phi^{(i)} - \Phi^{(i)}\,\nabla^{b} \bar{\Phi}^{(i)}\right)~,
\end{equation}
to define the conserved Noether charge associated to the system:
\begin{equation}
N^{(i)} \equiv \int dx^{3}\sqrt{-g}\,g^{0a}J^{(i)}_{a}~, 
\label{noe} 
\end{equation}
which can be interpreted as the number of bosonic particles in the star~\cite{liebpa}.

We consider two extreme cases, modeling very different behavior of the scalar field cores, to study the diverse phenomenological dynamics. In the first case, which we study in depth, each DM core forms an independent Bose-Einstein condensate and is assumed to interact with the other DM and with the fermionic component only through gravity, which seems a rather plausible approximation. Consequently, the bosonic matter is effectively modeled in our simulations by using two independent complex scalar fields, $\Phi^{(1)}$ and $\Phi^{(2)}$, corresponding to the DM cores inside each star, which we call \textit{non-interacting scalar field model} (NISF). 

The second case, considered mainly for comparison purposes, allows the bosonic matter to interact not only through gravity but also by means of scalar field interactions. In this particular case, that we call \textit{interacting scalar field model} (ISF), both DM cores are described 
by using the same single complex scalar field, thus directly coupling the two bosonic components via the KG equation.

Note that the merger of DM cores with no fermionic matter, so-called boson stars, have been previously studied both in the NISF~\cite{bezpalen2} and in the ISF cases~\cite{bezpalen,palenpani}. Here we focus on FBSs only.

\subsubsection{Fermionic matter}

We describe the fermionic component through the energy-momentum tensor for a perfect fluid, given by
\begin{equation}
T^{M}_{ab}=[\rho(1+\epsilon)+P] u_{a}u_{b} + P\,g_{ab}~,
\label{matter_tensor}
\end{equation}
where $\rho$ is the rest mass density of the fluid, $\epsilon$ its specific internal energy, $P$ its pressure, and $u^{a}$ is the velocity four-vector. The equations of motion are given by 
\begin{eqnarray}
\nabla_{a}(\rho u^{a}) &=& 0~,\label{matter_FB1}\\
\nabla^{a}T^{M}_{ab} &=& 0~,\label{matter_FB2}
\end{eqnarray}
which ensures the conservation of the rest mass and the energy-momentum, respectively.

\subsection{Evolution formalism}
The EKGH covariant equations can be written as an evolution formalism by performing the $3+1$ decomposition~\cite{CCC,gour}. The line element can be decomposed as
\begin{equation}
  ds^2 = - \alpha^2 \, dt^2 + \gamma_{ij} \bigl( dx^i + \beta^i dt \bigr) \bigl( dx^j + \beta^j dt \bigr)~, 
\label{3+1decom}  
\end{equation}
where $\alpha$ is the lapse function, $\beta^{i}$ is the shift vector, and $\gamma_{ij}$ is the induced metric on each spatial foliation
. In this foliation, the normal to the hypersurfaces 
can be defined as $n_{a}=(-\alpha,0)$ and the extrinsic curvature as $K_{ij} \equiv  -\frac{1}{2}\mathcal{L}_{n}\gamma_{ij}$,  where $\mathcal{L}_{n}$ is the Lie derivative along  $n^{a}$. 

Subsequently, a conformal decomposition is applied to the evolved fields, which basically consists of performing a conformal transformation to the metric
and the extrinsic curvature, i.e., $\gamma_{ij}$ into $\tilde{\gamma}_{ij}$ with unit determinant and $K_{ij}$ into a trace $trK$ and a trace-less part $\tilde{A}_{ij}$. This transformation leads to two new constraints, which can also be enforced dynamically by including additional damping terms to the evolution equations~\cite{bezpalen}.  The final set of evolution fields with the gauge conditions for the lapse and shift
 can be found in~\cite{bezpalen,simf3}. 

\subsubsection{Bosonic matter evolution: Klein-Gordon equations}

The KG equations~\eqref{FB_EKG1} can also be written as a time evolution system by performing the $3+1$ decomposition. First, we define 
\begin{equation}
\Pi^{(i)} \equiv
-\frac{1}{\alpha}\left(\partial_{t}-\beta^{k}\partial_{k}\right) \Phi^{(i)}~,
\end{equation}
as a new evolved field. In terms of the 3+1 quantities, the evolution equations for each complex scalar field can be written as
\begin{eqnarray}
  \partial_t \Phi^{(i)} &=& \beta^k \partial_k \Phi^{(i)} - \alpha \Pi^{(i)}~,
\\
  \partial_t \Pi^{(i)} &=& \beta^k \partial_k \Pi^{(i)} \nonumber
   +  \alpha \left[ -\gamma^{ij} \nabla_i \nabla_j \Phi^{(i)} + \Pi^{(i)} \,tr K   \right.\nonumber \\
   &+& \left.\frac{dV^{(i)}}{d \left|\Phi^{(i)}\right|^2} \Phi^{(i)} \right]  - \gamma^{ij} \nabla_i \Phi^{(i)} \nabla_j \alpha~,
\end{eqnarray}
which are still generic for any self-potential.

\subsubsection{Fermionic matter: General Relativistic Hydrodynamics equations}

First of all, a $3+1$ decomposition to the four-velocity vector $u^{a}$ is applied by writing it down in terms of a parallel and orthogonal part to the vector $n^{a}$, namely
\begin{equation}
u^{a} = W(n^{a} + v^{a})~,
\end{equation}
where $W=-n_{a}u^{a}$ is the Lorentz factor and $v^{a}$ is the three-velocity vector, both of them measured by Eulerian observers. 

The General Relativistic Hydrodynamics equations (GRHD) evolution equations are usually written in flux-conservative form, namely 
\begin{equation}
\partial_{t} \textbf{u}
+\partial_{k}F^{k}(\textbf{u}) = S(\textbf{u})~,
\end{equation} 
which allow to use numerical methods to deal with the inherent shocks appearing due to the non-linearities of the equations. Here $\textbf{u}$ is a vector of conserved fields, which will be defined below. Within this framework, the equation of continuity $\eqref{matter_FB1}$ and the energy-momentum conservation \eqref{matter_FB2} read:
\begin{eqnarray}
&&\partial_{t}(\sqrt{\gamma} D) +  \partial_{k}\left[\sqrt{\gamma}(-\beta^{k} + \alpha\,v^{k})D\right]=0~,\label{D_eq}\\
&&\partial_{t}(\sqrt{\gamma} \tau) + \partial_{k}\left[\sqrt{\gamma}(-\beta^{k}\tau + \alpha\,(S^{k}-\tau))\right]\nonumber\\
&&\qquad=\,\sqrt{\gamma}(\alpha\,S^{ij}K_{ij}-S^{j}\partial_{j}\alpha)~, \label{tau_eq}\\
&&\partial_{t}(\sqrt{\gamma}S_{i}) + \partial_{k}\left[\sqrt{\gamma}(-\beta^{k}S_{i}+\alpha\,S^{k}_{i})\right]\nonumber\\
&&\qquad=\,\sqrt{\gamma}\left(\alpha\,\Gamma^{j}_{ik}S^{k}_{j}+S_{j}\partial_{i}\beta^{j}-(\tau+D)\partial_{i}\alpha\right)~,\label{Si_eq}
\end{eqnarray}
where $\gamma=\det(\gamma_{ij}).$ The evolved conserved variables $\textbf{u}=\{D,\tau,S_{i}\}$ are proportional to the rest-mass density measured by Eulerian observers $D$, the energy density $\tau$ (i.e., without the mass density) and the momentum density $S_{i},$ which are defined as
\begin{eqnarray}
D&:=& - n_{a}\rho\,u^{a}=\rho W~,\label{DD}\\
\tau&:=& n^{a}n^{b}T^{M}_{ab} = h W^{2} - P - D,\\
S_{i}&:=& -\gamma_{i}^{a}n^{b}T^{M}_{ab}= hW^{2}v_{i}~,
\end{eqnarray}
where $h = \rho(1+\epsilon) + P$ is the enthalpy and $W = 1/\sqrt{1 - \gamma^{ij} v_i v_j}$ is the Lorentz factor in terms of the three-velocity vector. Finally, $S_{ij}$ is the spatial projection of the energy-momentum tensor $T^{M}_{ab},$ namely 
\begin{equation}
~S_{ij} :=\gamma_{i}^{c}\gamma_{j}^{d}T^{M}_{cd}=hW^{2}v_{i}v_{j} + P\,\gamma_{ij}~.
\end{equation} 
Furthermore, to recover the physical or \textit{primitives fields} $(\rho,\epsilon,P,v_{i})$, required to perform the evolution,  an  equation of state (EoS) must to be imposed. During the evolution we employ an ideal-gas EoS, $P = (\Gamma - 1)\rho\epsilon$, where $\Gamma$ is the adiabatic index and it assumed to be a constant, which is able to capture the fluid heating due to strong shocks produced in the merger stage~\cite{CCC,SHIBATAbook}. The transformation from conserved to primitive fields involves non-linear equations which, in general, need to be solved numerically~\cite{Neilsen_2006}.

\subsection{Numerical setup and analysis}\label{numeri_fb}
We adopt finite difference schemes, based on the method of lines~\cite{1995tpdm.book.....G}, on a regular Cartesian grid. A fourth-order-accurate spatial centered discretization (satisfying the summation-by-parts rule) is used for  Einstein equations~\cite{cal}. The relativistic hydrodynamics equations are discretized using High-Resolution-Shock-Capturing method based on the Harten-Lax-van Leer-Einfeldt flux formula~\cite{Harten:1983,Toro:1997} with piecewise parabolic reconstruction~\cite{Colella:1984,MARTI19961,COLELLA20087069}. Finally, a third-order-accurate Runge-Kutta scheme is used to integrate the equations in time~\cite{ander}.

To ensure  sufficient resolution, we employ adaptive mesh refinement (AMR) via the \textsc{had} computational infrastructure that provides distributed, Berger-Oliger style AMR~\cite{had,lieb} with full sub-cycling in time, together with an improved treatment of artificial boundaries \cite{lehner}. We adopt a Courant parameter, defined by the ratio between the timestep and the grid size, $\lambda_c \approx 0.25$ such that $\Delta t_{l} = \lambda_c \, \Delta x_{l}$ on each refinement level $l$ to guarantee that the Courant-Friedrichs-Levy condition is satisfied. Our numerical implementation has been tested with several benchmark problems dealing with Einstein-KG equations~\cite{pale1,pale2,palenpani,bezpalen,bezpalen2} and GRHD equations~\cite{PhysRevD.92.044045,PhysRevLett.100.191101,PhysRevD.77.024006}.

Finally, the gravitational radiation can be  calculated by computing the Newman-Penrose scalar $\Psi_{4}$ and expanding it in a basis of spin-weighted spherical harmonics (with spin weight $s=-2$)~\cite{rezbish,brugman}, namely
\begin{equation}
r \Psi_4 (t,r,\theta,\phi) = \sum_{l,m} \Psi_4^{l,m} (t,r) \, {}^{-2}Y_{l,m} (\theta,\phi)~,
\label{eq:psi4}
\end{equation}
where $\Psi_4^{l,m} (t,r)=\int_{S^{2}}\Psi_{4}{}^{-2}Y_{l,m}d\Omega$.

\section{Initial data}\label{id_fb}	
Here, we explain in detail how to construct initial data for equilibrium configuration of non-rotating isolated FBSs by using the methodology explained in Ref.~\cite{suspalen}. We also summarize the procedure to construct initial data for binary FBSs.

\subsection{Isolated Fermion-Boson star}
The equilibrium configuration equations for a single FBS can be obtained by combining the procedures to obtain isolated NS and boson stars solutions~\cite{CCC}. Let us start by assuming a static metric given by the line element in Schwarzschild coordinates (polar-areal coordinates~\cite{polar}):
\begin{equation}
ds^{2}=-\alpha^{2}(\tilde{r})dt^{2}+a^{2}(\tilde{r})d\tilde{r}^{2}+\tilde{r}^{2}d\Omega^{2}~.
\label{polar_metric}
\end{equation}
We also impose the static fluid condition $v^{i}=0$ and an harmonic time dependence ansatz for the scalar field
\begin{equation}
   \Phi(t,\tilde{r}) = \phi_{0}(\tilde{r})\,e^{-i\omega t}~,
\label{scf}
\end{equation}
where $\omega$ is a real frequency and $\phi_{0}(\tilde{r})$ is a real-value spatial function. Under these conditions, the EKGH system lead to the following system of ordinary differential equations (ODE):
\begin{eqnarray}
\partial_{\tilde{r}}a &=& -\frac{a}{2\tilde{r}}\left( a^{2} - 1\right) +  4\pi\,\tilde{r}\,a^{3}\tau~, \label{a}\\
\partial_{\tilde{r}}\alpha &=& -\frac{\alpha}{2\tilde{r}}( 1 - a^{2} ) + 4\pi\,\tilde{r}\,\alpha\,a^{2}\,S_{\tilde{r}}^{\tilde{r}}~,\\
\partial_{\tilde{r}}\phi_{0} &=& \zeta~,  \\
\partial_{\tilde{r}}\zeta &=& -\left( 1 + a^{2} + 4\pi\,\tilde{r}^{2}\,a^{2}\right)\left(S^{\tilde{r}}_{\tilde{r}} - \tau\right)\frac{\zeta}{\tilde{r}} \nonumber\\
&& - \left(\left(\frac{\omega}{\alpha}\right)^2 - \frac{dV}{d \left|\Phi\right|^2} \right)a^{2}\phi_{0}~,\\
\partial_{\tilde{r}}P &=& -\left(\rho(1+\epsilon)+P\right)\frac{\partial_{\tilde{r}}\alpha}{\alpha}\label{P}~,
\end{eqnarray}
where 
\begin{eqnarray}
\tau &=& \left(\frac{\omega\,\phi_{0}}{\alpha}\right)^{2} +\left(\frac{\zeta}{a}\right)^{2} +V(|\Phi^{2}|)+ h - P~,\\
S^{\tilde{r}}_{\tilde{r}} &=& \ \left(\frac{\omega\,\phi_{0}}{\alpha}\right)^{2} +\left(\frac{\zeta}{a}\right)^{2} - V(|\Phi^{2}|) + P~,\\
S^{\theta}_{\theta} &=& -\left(\frac{\omega\,\phi_{0}}{\alpha}\right)^{2} +\left(\frac{\zeta}{a}\right)^{2} - V(|\Phi^{2}|)+ P~. 
\end{eqnarray}
We adopt a polytropic EoS $P=\kappa \rho^{\Gamma}$, which is a reasonable approximation for cold NSs, considering the general purpose of our study, combined with the massive potential given by Eq.~\eqref{potential}.
The above system can be solved numerically by using boundary conditions guarantying regularity at the origin and asymptotic flatness, namely
\begin{eqnarray*}
a(0) &=& 1,~\alpha(0) =1~,\\
\phi_{0}(0) &=&\phi_{c},~\zeta(0)=0,~P(0)=\kappa\rho_{c}^{\Gamma}~,\\
\lim_{\tilde{r}\to\infty}\phi_{0}(\tilde{r}) &=& 0,~\lim_{\tilde{r}\to\infty}\alpha(\tilde{r}) = \frac{1}{a(\tilde{r})}, \lim_{\tilde{r}\to\infty}P(\tilde{r}) = 0~,
\end{eqnarray*}
where $\phi_{c}$ is the central value of scalar field and $\rho_{c}$ the central value of rest-mass density.  Notice that the final ODE system constitutes an eigenvalue problem 
for $\omega$ as a function of $\{\phi_{c},\rho_{c}\}$. A shooting method can be used in order to integrate the system~\eqref{a}-\eqref{P} from $\tilde{r}=0$ towards the outer boundary. 

In addition, we can add two global conserved quantities to help on the characterization of solutions, namely the fermionic rest-mass and the bosonic rest-mass. The profiles of these quantities, contained within a radius $\tilde{r}$, for the equilibrium solutions, are given by the following differential equations
\begin{eqnarray}
\frac{\partial M_{B}}{\partial\tilde{r}} & =& 8\pi \mu \omega\,\frac{\phi^{2}_{0}\,a\,\tilde{r}^{2} }{\alpha}~,\\
\frac{\partial M_{F}}{\partial\tilde{r}} & =& 4\pi\rho\,\tilde{r}^{2}~,
\end{eqnarray}    
with boundary conditions $M_{B}(0)=M_{F}(0)=0$. Hereafter, we will indicate with $M_F$ and $M_B$ the total fermionic and bosonic masses. 

The radius of the bosonic component, denoted as $R_{B}$, is defined as the distance from the center at which $99\%$ of the bosonic mass is contained. The radius of the fermionic component, $R_{F}$, is instead defined as the radius where pressure vanishes, as usual in standard NSs.

Assuming a spherically symmetric solution, the Arnowitt-Deser-Misner (ADM) total mass of each FBSs can be computed as:
\begin{equation}
M=\lim_{\tilde{r}\to\infty}\frac{\tilde{r}}{2}\left(1-\frac{1}{\alpha(\tilde{r})^{2}}\right)~.
\end{equation}

Finally, after the equilibrium configuration is found, a coordinate transformation from polar to isotropic coordinates is performed.  Then, the solution can be easily written in Cartesian coordinates to perform our numerical 3D simulations~\cite{mundim}.

The equilibrium configurations  depend on two parameters: the central values of the scalar field, $\phi_{c}$, and of the rest-mass fermionic density $\rho_{c}$. By varying these parameters, together with the EoS and the potential of the scalar field, it is possible to find star solutions composed mostly either by fermions (i.e., $M_F\gg M_B$) or bosons (i.e., $M_B\gg M_F$). 
Solutions can then be characterized by the boson-to-fermion ratio:
\begin{equation}
   Y_{B}=\frac{M_{B}}{M_F}~.
\label{nbnf}
\end{equation}
For a fixed value of the total mass of the star, the mass of bosons $M_{B}$ grows as $\phi_{c}$ (or $\rho_{c}$) increases, reaching a maximum, which is not shown in our figures, and decreasing afterwards. The mass of fermions $M_{F}$ follows the complementary behavior to $M_{B},$ i.e., it decreases for increasing $\phi_c$, reaches a minimum, and then increases. It is worth stressing that the stability criteria for a FBS is not trivial, since it depends on two parameters, $\{\phi_{c},\rho_{c}\}$ (see \cite{suspalen} and references therein).

Here we are going to consider a polytropic EoS with $\Gamma=2.5$ and $\kappa=8980$ (in geometric units), leading to configurations with fermionic mass and radius in the range of typical NS. The parameter $\mu = m_{b}c/\hbar$ has dimension of inverse of length in geometric units, where $m_{b}$ is the boson mass (see for instance~\cite{Dietrich_2018}).
We shall set $\mu=1$ in our simulations, which  allows to have the same order of magnitude for $R_B$ and $R_F$, and at the same time construct stable stars dominated by the fermionic mass but with a non-negligible bosonic component $Y_B\lesssim 0.1$. In order to transform $\mu$ into cgs units one needs to multiply by $c^{2}/GM_{\odot}$, meaning that $\mu=1$ is equivalent to have a boson mass of $m_b=8.3\times 10^{-10}\,eV/c^{2}.$

\begin{figure}
\centering
\includegraphics[width=1.05\linewidth]{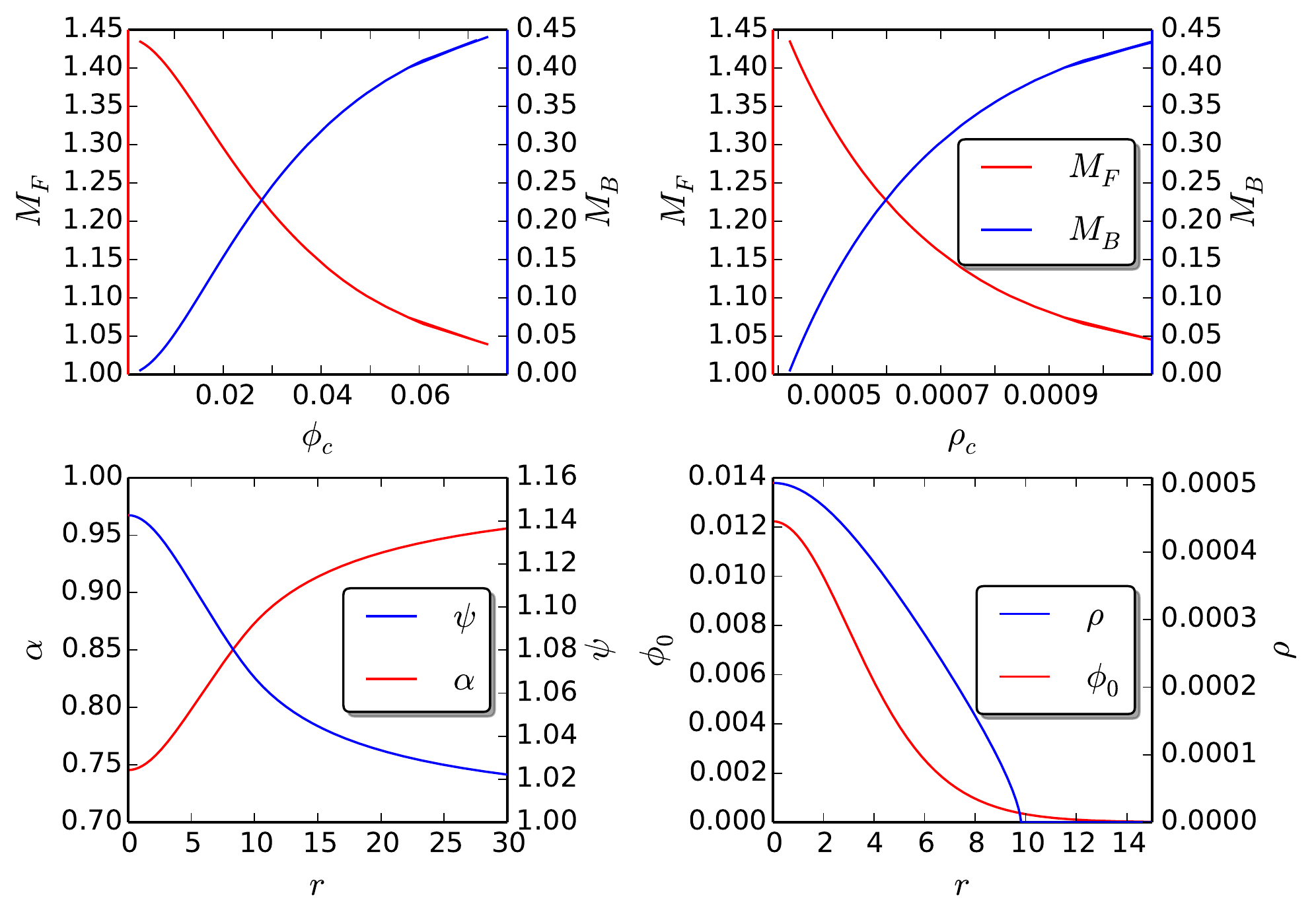}
\caption{{\em Initial data of an isolated FBS.} (Top) The mass of fermions $M_{F}$ and bosons $M_{B}$ for the equilibrium configurations with a fixed total ADM mass $M=1.35$, as a function of the central value of scalar field $\phi_{c}$ (left) and  rest-mass density $\rho_{c}$ (right). (Bottom) The metric component radial profiles $\alpha(r)$ and $\psi(r)$ (left), together with the ones for scalar field $\phi_{0}(r)$ and density $\rho(r)$ (right), for a specific stable equilibrium with boson-to-fermion ratio $Y_{B}=10\%$.}
\label{twin}
\end{figure}

By considering a fixed ADM mass $M=1.35M_{\odot}$ we can find a family of equilibrium configurations. The behavior of $M_{F}$ and $M_{B}$ explained above, for this particular family, is shown in the top panel of Fig.~\ref{twin}. In the bottom panel of Fig.~\ref{twin}, we show the profiles of $\phi_{0}(r),$ $\rho(r)$, $\alpha(r)$ and $\psi(r)$ for an illustrative solution obtained for the choice $\phi_{c}=1.223\times 10^{-2}$ and $\rho_{c}=5.0244\times 10^{-4}$, leading to a stable equilibrium configuration with $Y_{B}=10\%$, compactness $C= M/R_{F} = 0.12$ and $R_{F}=11.2$. Note that, due to the chosen EoS, our stars have larger fermionic radii than what expected from standard NSs (and from their observational constraints), but it is not crucial for studying the influence of DM on the dynamics at a qualitative level.

\begin{figure}
	\centering
	\includegraphics[width=1.0\linewidth]{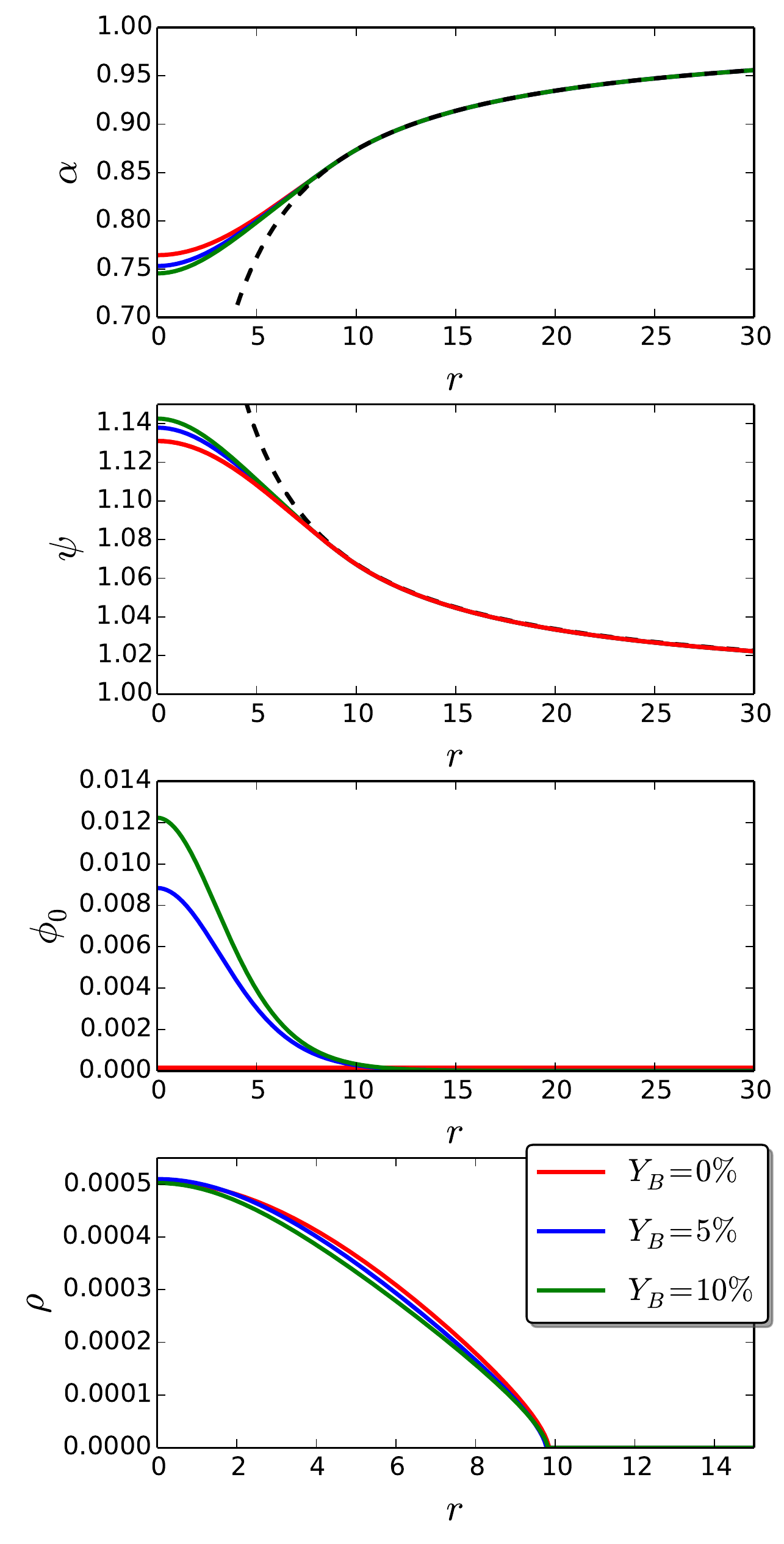}
	\caption{{\em Initial configuration for each FBS considered in the binary system.} The radial profiles of the metric components $\alpha(r)$ and $\psi(r)$ (top and second panel), the radial scalar field $\phi_{0}(r)$ (third panel) and the radial density profiles (bottom panel) respectively, for the different boson-to-fermion ratios $Y_B$. The dashed black lines show the Schwarzchild solution with the same mass of the FBS.}   
	\label{static_fb}
\end{figure}

In the next section, we will consider binary FBSs obtained as a superposition of three different isolated FBS configurations belonging to the stable branch. Such configurations consist of the same individual ADM mass of $M=1.35M_{\odot}$ and compactness $C=0.12$, but different boson-to-fermion ratios $Y_{B}=\{0\%,5\%,10\%\}$.
These  equilibrium configurations are constructed by using a polytropic EoS with $\Gamma=2.5$, but varying the polytropic constant $\kappa$ to achieve solutions with the same ADM mass and compactness. The radial profiles of the metric components $\alpha(r)$ and $\psi(r)$, the scalar field $\phi_{0}(r)$ and the density $\rho(r)$, for these three configurations, are displayed in  Fig.~\ref{static_fb}. Obviously, the number of boson increases with the scalar field. 

\begin{figure}
\centering
\includegraphics[width=1.0\linewidth]{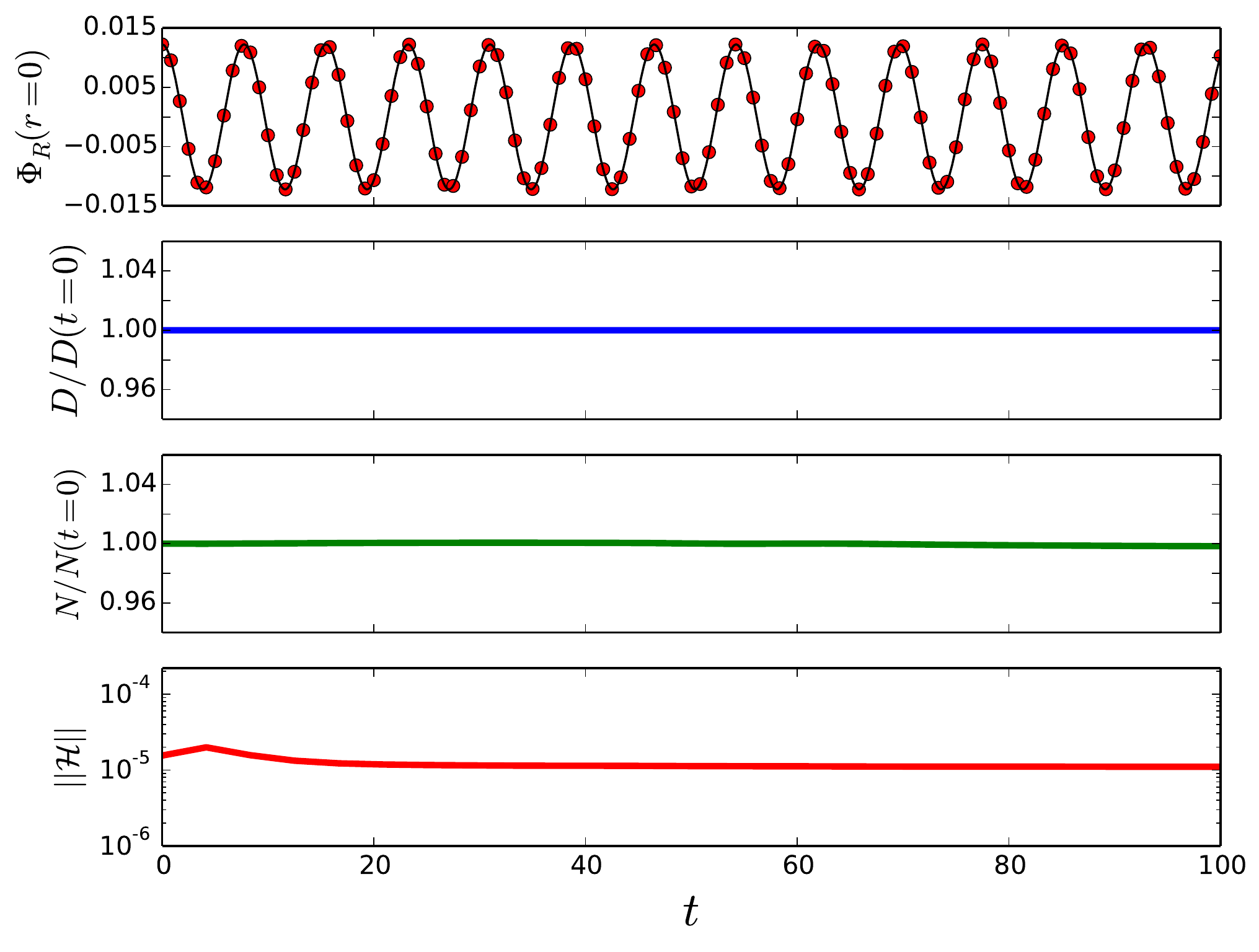}
\caption{{\em Evolution of an isolated FBS.} (Top) Evolution of the real part of $\Phi$ at $r=0$ for the stable configuration with ADM mass $M=1.35$ and $Y_{B}=10\%$. The numerical solution (red circles) is in very good agreement with its analytical expected value (solid black), given by $\phi_{0}(r=0)\cos(\omega t)$ with $\omega=1.0878$. (Second and third row) Evolution of the integrated rest-mass density $D$ and the Noether charge $N$, showing that they remain roughly constant during the evolution, as expected for a stable configuration. (Bottom) $L_{2}$ norm of the Hamiltonian $\mathcal{H}$ as a function of time, showing that the constraints are small and remains under control during the simulation.}
\label{phi_ham_D}
\end{figure}

The stability of the isolated FBS configuration
represented in Fig.~\ref{twin} can be tested by evolving them, solving the EKGH equations described above. We set a domain $[-100,100]^{3}$ with radiative boundary conditions, using $120$ grid point in each direction and four refinement levels, such that the finest one has a resolution of $\Delta x_{3} = 0.21$. The dynamical evolution of some relevant stability indicators is displayed in Fig.~\ref{phi_ham_D}. In particular, the real part of the scalar field at the center $\Phi_{R}(t,r=0)$ is displayed in the top panel, and compared to the expected analytical behavior $\phi_{0}(r=0)\cos(\omega t)$, showing a perfect agreement. The spatial integral of globally conserved quantities, namely the rest-mass density $D$  and the Noether charge $N$, are showed in the second and third panel. These quantities have been rescaled by their initial values. Notice that they remain roughly constant during the evolution, confirming that the initial equilibrium configuration is stable. Finally, the $L_{2}$ of the Hamiltonian constraint is displayed in the bottom panel, showing that the violation of this constrain remains under control during the evolution and it is comparable to its initial value, which is given mainly by discretization errors.

\subsection{Binary Fermion-Boson stars}
Initial data for binary FBSs can be constructed by a superposition of two boosted isolated FBS solutions. In a previous work~\cite{bezpalen},  we have explained in detail how to boost a static spherically symmetric solution and the scalar fields quantities with a velocity $v$ along the $x$-axis. Here, we extend this procedure to include also the hydrodynamical fields. We start by performing a Lorentz transformation to the four-velocity vector $u^{a}$, namely 
\begin{eqnarray}
\tilde{u}_{a} = \Lambda^{b}_{a}u_{b}~,
\end{eqnarray}
where the matrix related to the transformation has the following form
\begin{eqnarray}
\Lambda^{b}_{a} =\begin{pmatrix}
\Gamma & -\Gamma\,v & 0 & 0 \\ 
-\Gamma\,v & \Gamma & 0 & 0 \\ 
0 & 0 & 1 & 0 \\ 
0 & 0 & 0 & 1
\end{pmatrix}~,
\end{eqnarray}
being $\Gamma$ the general relativistic Lorentz factor related to the transformation. Therefore, at time $t=0,$ we obtain
\begin{eqnarray}
\tilde{u}_{t} = -\Gamma\,\alpha~,~~ \tilde{u}_{x} =  \Gamma\,v\,\alpha~,~~ \tilde{u}_{y} = \tilde{u}_{z} = 0~.
\end{eqnarray}
Finally, the final expression for the hydrodynamics fields  of the boosted star, evaluated at $t=0$ are: 
\begin{eqnarray}
\tilde{\rho}=\rho~,~\tilde{v}_{x} = \frac{\tilde{u}_{x}}{\tilde{W}}~,~~\tilde{v}_{y} =  \tilde{v}_{z} = 0~,
\end{eqnarray}
where $\tilde{W}^{2}= \tilde{u}_{i}\tilde{u}^{i} + 1.$ 

The method to construct initial data for FBS binaries can be summarized as follows:
\begin{itemize}
\item the solution of each isolated FBS $(i)$ is written in Cartesian coordinates:   $\{g^{(i)}_{ab}(x,y,z),\Phi^{(i)}(x,y,z),\rho^{(i)}(x,y,z),v_{j}^{(i)}(x,y,z)\}$.

\item the spacetime and the hydrodynamics fields of binary FBS are obtained by a superposition of the solution of two identical isolated FBS, centered at positions $(0,\pm y_{c},0)$ and with a boost $\pm v_x$ along the $x$-direction, namely 
\begin{eqnarray}
	g_{ab} &=& g^{(1)}_{ab}(x,y-y_{c},z;+v_{x})\nonumber\\
	&+& \,g^{(2)}_{ab}(x,y+y_{c},z;-v_{x}) - \eta_{ab}~,\\
	\rho &=& \rho^{(1)}(x,y-y_{c},z;+v_{x})\nonumber \\
	&+& \,\rho^{(2)}(x,y+y_{c},z;-v_{x})~, \\
	v_{j} &=& v^{(1)}_{j}(x,y-y_{c},z;+v_{x})\nonumber\\
	&+& \,v^{(2)}_{j}(x,y+y_{c},z;-v_{x})~,
\end{eqnarray}
where $\eta_{ab}$ is the Minkowski metric in Cartesian coordinates.

\item as we explained in the previous section, we are interested on modeling FBSs binary systems in two different scenarios, where each scalar field is directly coupled (ISF) or not (NISF) to the other. The scalar fields are then initially given by the equilibrium solution for an isolated FBS, $\Phi^{(is)}$, centered at each fermion density maximum location:
\begin{itemize}
\item[(a)]NISF, i.e. two scalar fields are considered:
\begin{eqnarray}
\Phi^{(1)} &=& \Phi^{(is)}(x,y-y_{c},z;+v_{x})~,\\
\Phi^{(2)} &=& \Phi^{(is)}(x,y+y_{c},z;-v_{x})~.
\end{eqnarray}
\item[(b)]ISF, i.e. one scalar field is considered:
\begin{eqnarray}
\Phi &=& \Phi^{(is)}(x,y-y_{c},z;+v_{x})\nonumber \\
&+& \,\Phi^{(is)}(x,y+y_{c},z;-v_{x})~.
\end{eqnarray}
\end{itemize}
\end{itemize}
It should be stressed that a fine-tuning of the initial orbital velocity is required to set the binary in a quasi-circular orbit. Notice that the construction by a mere superposition does not satisfy the energy and momentum constraints due to the non-linear character of Einstein's equations. Nevertheless, the CCZ4 formalism used enforces dynamically an exponential decay of these constraint violations, see e.g. Fig.~10 in Ref.~\cite{bezpalen}.

The characteristics of our isolated FBS models, used to construct the binary systems, are summarized in Table~\ref{fb_conf}: central value of the density $\rho_{c}$, central value of the scalar field $\phi_{c}$, polytropic constant $\kappa$, angular frequency of the scalar field phase in the complex plane $\omega$, boson and fermion radii and mass of the star. We also include two quantities related to the coalescence: the merger time, defined as the time when the maximum of the norm of $\Psi^{2,2}_{4}$ is produced (peak of GW emission), and the frequency of the dominant peak in the Fourier spectral power distribution, calculated as the Fourier transform of $\Re(r\Psi^{2,2}_{4})$ during the post-merger stage. 
\begin{table*}
\begin{tabular}{c||c||cccc||cccc||cc}
 $Y_{B}$  & Model & $\rho_{c}$  & $\phi_{c}$ & $\kappa$ & $\omega$ & $R_B$ & ${R_F}$  & $M_{B}$ & $M_{F}$ & $t_{\rm merger}$ & $f_{\rm peak}$[kHz]
 \\ \hline\hline 
0  & NS & 5.0525$\times 10^{-4}$  & 0.0 &  7405 & 0  & 0 & 11.23 & 0.0 & 1.44 & 1650 &  1.62\\
5  & NISF & 5.0989$\times 10^{-4}$  & 8.838$\times 10^{-3}$   & 8136 & 0.814704559507  & 10.37 & 11.16 & 0.0721 & 1.37 & 1626 & 1.81\\
10 & NISF& 5.0244$\times 10^{-4}$  & 1.223$\times 10^{-2}$   & 8980 & 0.811068278806  & 10.15 & 11.20 & 0.1262 & 1.32 & 1606 & 1.87\\
10 & ISF & 5.0244$\times 10^{-4}$  & 1.223$\times 10^{-2}$   & 8980 & 0.811068278806  & 10.15 & 11.20 & 0.1262 & 1.32 & 1616 & 1.93 
\end{tabular}
\caption{{\em Summary of the binary FBS configurations.} The table shows: boson-to-fermion ratio $Y_{B},$ central value of the scalar field, polytropic constant $\kappa$, angular frequency of the phase of $\Phi$ in the complex plane, bosonic radius (containing $99\%$ of the Noether charge), fermionic radius (i.e, the radius where the fluid pressure vanishes), bosonic and fermionic masses. All models have ADM mass $M=1.35$ and compactness $C=0.12$. The last two columns are related to the simulation results: merger time, defined as the one corresponding to the maximum of the norm of the $\Psi^{2,2}_{4}$, and frequency of the dominant peak in the power spectral density of the $\Psi^{2,2}_{4}$, evaluated during the post-merger stage.}
\label{fb_conf}
\end{table*}

\section{Coalescence of Fermion-Boson Stars}\label{coales_fb}
In the present section we study the dynamics of the coalescence. Our simulations are performed in a domain $[-480,480]^3$ with a coarse resolution of $\Delta x_{0}=6.8$. There are $6$ levels of refinement, the last of which has $\Delta x_{5}=0.21$ and is designed to cover only the stars before and after the merger. FBSs are initially centered at $(0,\pm 16,0)$ and have a boost velocity $v_{x}=\pm 0.173$, leading to a binary system in a tight quasi-circular orbit. 

We will focus on the dynamics of the cases shown in Table~\ref{fb_conf}: a standard NS, two NISF and one ISF. For numerical convergence purposes, we have performed numerical simulation for the case NISF $Y_{B}=10\%$ with higher resolution, finding the same qualitative and quantitative results shown below (both in dynamics and in the power spectra).

Besides the fermionic and bosonic density distributions, we will analyze in detail the amplitude and the power spectral density of the gravitational radiation emitted during the coalescence, which is encoded in the Newman-Penrose scalar $\Psi_{4}$. This scalar is numerically integrated over a spherical surface at $R_{ext} = 120$, already located in the wave-zone.

\subsection{Dynamics}
\begin{figure*}
	\centering 
	\includegraphics[width=1.0\linewidth]{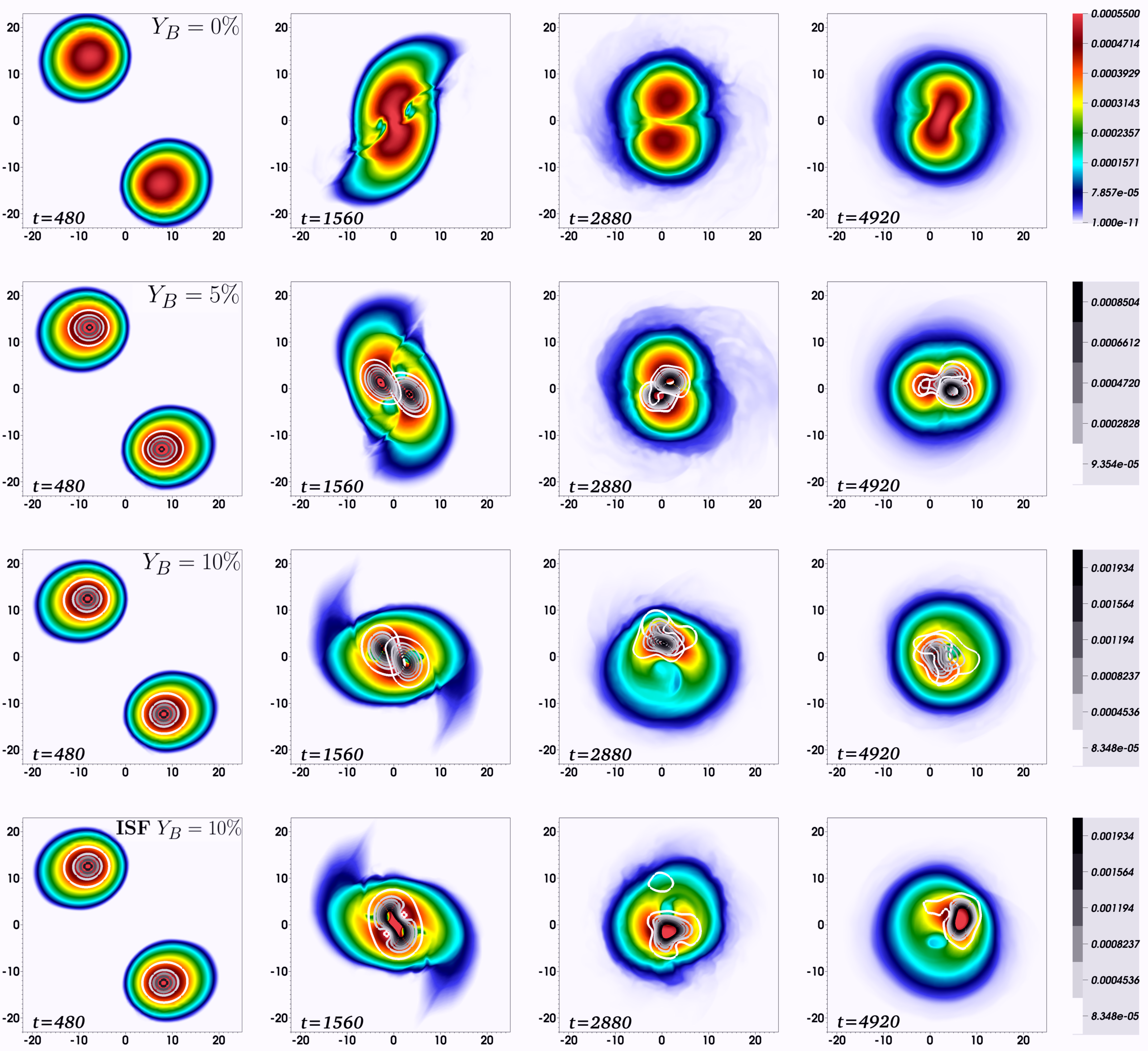}
	\caption{{\em{Dynamics of binary FBSs coalescence.}} Rest-mass densities for the fermionic components are represented in colors, while Noether charge densities are displayed in white-to-black contours, in the equatorial plane $(z=0)$, at different representatives times. The rows correspond  to the cases (from top to bottom) NS, NISF with $Y_{B}=\{5\%,10\%\}$, and ISF with $Y_{B}=10\%$. The first column illustrates a time in the early inspiral, the second one is roughly at merger time, the third one is during the post-merger stage and the fourth one at the end of our simulation.}
	\label{snapshot_fb}
\end{figure*}

The main dynamics can be inferred from the snapshots on the equatorial plane of the rest-mass and Noether charge densities for all these cases, at relevant times of the coalescence, displayed in Fig.~\ref{snapshot_fb}. 
In all cases, with the given initial separation, the stars complete two full orbits before colliding. During the inspiral stage (leftmost column), both the fermionic and the bosonic components of each star follow the same trajectory, roughly maintaining individually the initial equilibrium structure, except for some minor oscillations in each star due to the transient initial data adjustment. We do not believe that these affect the results, but to be sure one should repeat the simulations with equilibrium, constraint satisfying initial data. During this stage, the internal structure of the stars does not play an important role, and the presence of the bosonic component has a minor effect, as can be observed from the comparison with the NS case with the same ADM mass, i.e. $Y_B=0\%$.

At contact time $t \sim 1300$ from the beginning of the simulation, the fermionic components of both stars (indicated with colors) first touch (being their centers at a distance $\sim 2\,R_{F}$), and start merging into a single remnant (second column). When there is no bosonic component ($Y_B=0\%$), the newly formed object consists of a differentially rotating massive NS. This rotating remnant, with a shape dominated by a quadrupolar structure (third column), produces GWs mostly in the modes $l=|m|=2$,  see~\cite{doi:10.1146/annurev-astro-081913-040031} and reference therein. When $Y_B \ne 0\%$, the bosonic components (indicated with white-to-black contours) are gravitationally bound, but since they represent different scalar fields they do not merge into a single object (see the second column).  After the contact time, each bosonic core maintains roughly its shape while orbiting within the rotating remnant, eventually overlapping in space and forming a superposition of two coexisting boson cores. During the post-merger stage, in the case NISF with $Y_B=10\%$ (third row of Fig.~\ref{snapshot_fb}), the presence of different components in the system (i.e, two fermionic cores and two bosonic cores), makes the entire system more unstable due to the gravity coupling, exciting a relatively strong one-armed spiral instability~\cite{PhysRevD.92.121502,PhysRevD.89.044011,m=1} that breaks the quadrupolar structure of the system.

\begin{figure}[h!]
\centering
\includegraphics[width=0.9\linewidth]{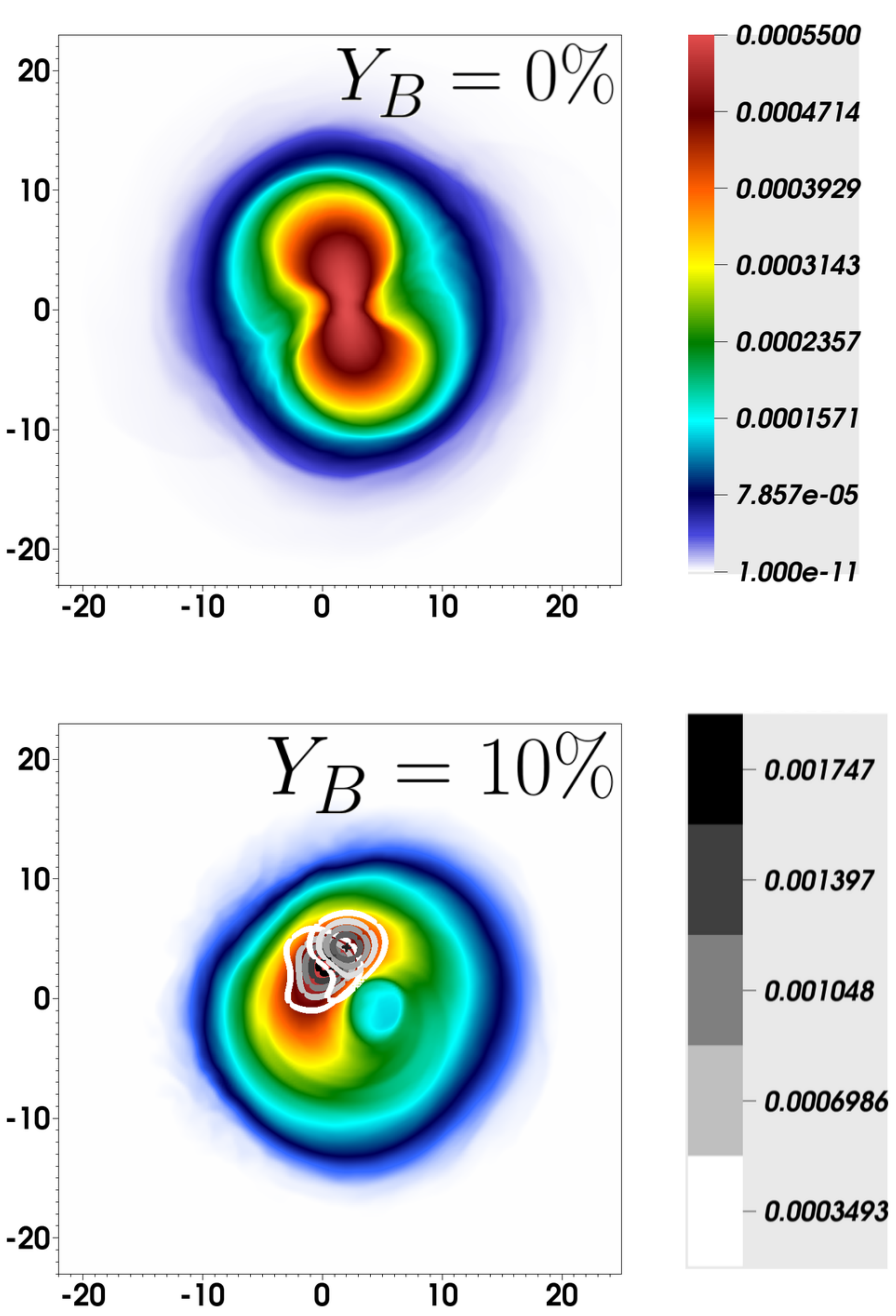}
\caption{ {\em{Remnant comparison}}. Rest-mass fermionic densities in the equatorial plane in the post-merger stage (i.e., roughly at $t-t_{merger}\simeq 2824$) of the $Y_{B}=0\%$ (top panel) and NISF with $Y_{B}=10\%$ models. The Noether charge densities are added as white-to-black contours in the case $Y_{B}=10\%$.}
\label{lacksim}
\end{figure}

Figure~\ref{lacksim} displays with more detail the differences on the density profile in the equatorial plane $(z=0)$ between the cases $Y_B=0\%$ and NISF with $Y_B=10\%$ in the post-merger phase, showing the symmetry breaking of the quadrupole structure when there is a significant amount of bosonic component inside the NS. While the fermionic over-density appears, the bosonic component clusters in the same region, too, and the fermionic and the bosonic component rotate together.  

This lack of quadrupolar symmetry is also found in the case NISF $Y_B=5\%$ (second row-fourth column of Fig.~\ref{snapshot_fb}), although the full development of the $|m|=1$ over-density cannot be seen as clearly within the time reached by our simulation. As it was discussed in~\cite{m=1}, this instability can develop gravitational radiation with a significant $l=2$, $|m|=1$ component at the orbital frequency of the cores, i.e., half the frequency of the corresponding $l=|m|=2$ mode, as it will be discussed in detail in the next section.

The dynamics during the inspiral of the case with ratio $Y_{B}=10\%$ with ISF is similar to the case NISF $Y_{B}=10\%$ (last row of Fig.~\ref{snapshot_fb}). Differences arise only after the merger, when the scalar field interactions play an important role by forming a single largely-perturbed bosonic core inside the NS remnant. In this case the two bosonic components actually merge, since they are described by the same scalar field. Nevertheless, the one-armed spiral instability is seen anyway, probably because it is triggered by the initial presence of four components, which are strongly coupled only two by two (fermionic cores between them, and bosonic components between them).

\subsection{Gravitational wave radiation}

\begin{figure}[h]
\centering
\includegraphics[width=1.02\linewidth]{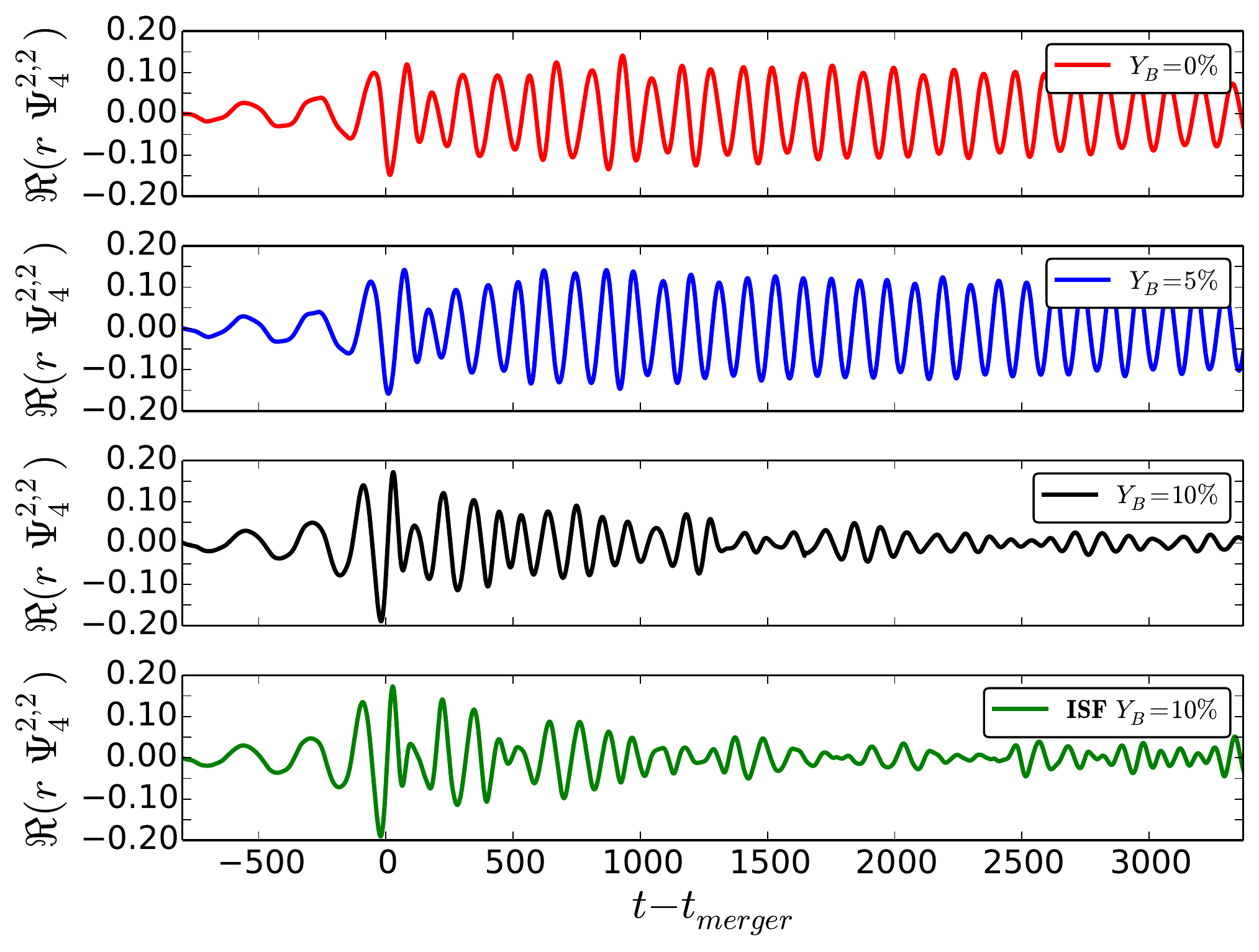}
\caption{{\em Gravitational waves.} The real part of the main $l=m=2$ mode of $\Psi_{4}$ describing the gravitational emission produced by binary FBSs coalescence, as a function of time on from the merger, for all models (top to bottom).}
\label{psi4_fb}
\end{figure}

An insight on the dynamics can be obtained by analyzing the GWs radiated by the system, which are described by the Newman-Penrose scalar $\Psi_{4}$, as a function of the time from merger, $t-t_{\rm merger}$. The amplitude (real part) of its main mode $l=m=2$ is displayed in Fig.~\ref{psi4_fb} for the four cases.
 
The gravitational radiation produced during the inspiral is roughly the same for all the cases, since the dynamics at this stage does not depend strongly on the inner structure of the stars, and the ADM mass is the same for all of them. As we saw above, qualitative and quantitative differences arise from the merger time on.

In all cases, strong quasi-periodic persistent oscillations are present soon after the merger, but notably their amplitude quickly decays in the case with the largest DM cores $Y_{B}=10\%$, both for NISF and ISF models. This notable qualitative difference could be possibly due to a quick redistribution of the density profile on the remnant, which becomes more axisymmetric in a shorter timescale due to the non-linear multi-body interaction between the two fermionic and the two DM components. Simultaneously, energy is transfered from the $|m|=2$ to the $|m|=1$ modes, probably through the one-arm instability~\cite{PhysRevD.93.024011}. The trigger to develop it, in our case, is the interaction between different components, which are coupled mostly or only by gravity.

\begin{figure}[h!]
\centering
\includegraphics[width=1.0\linewidth]{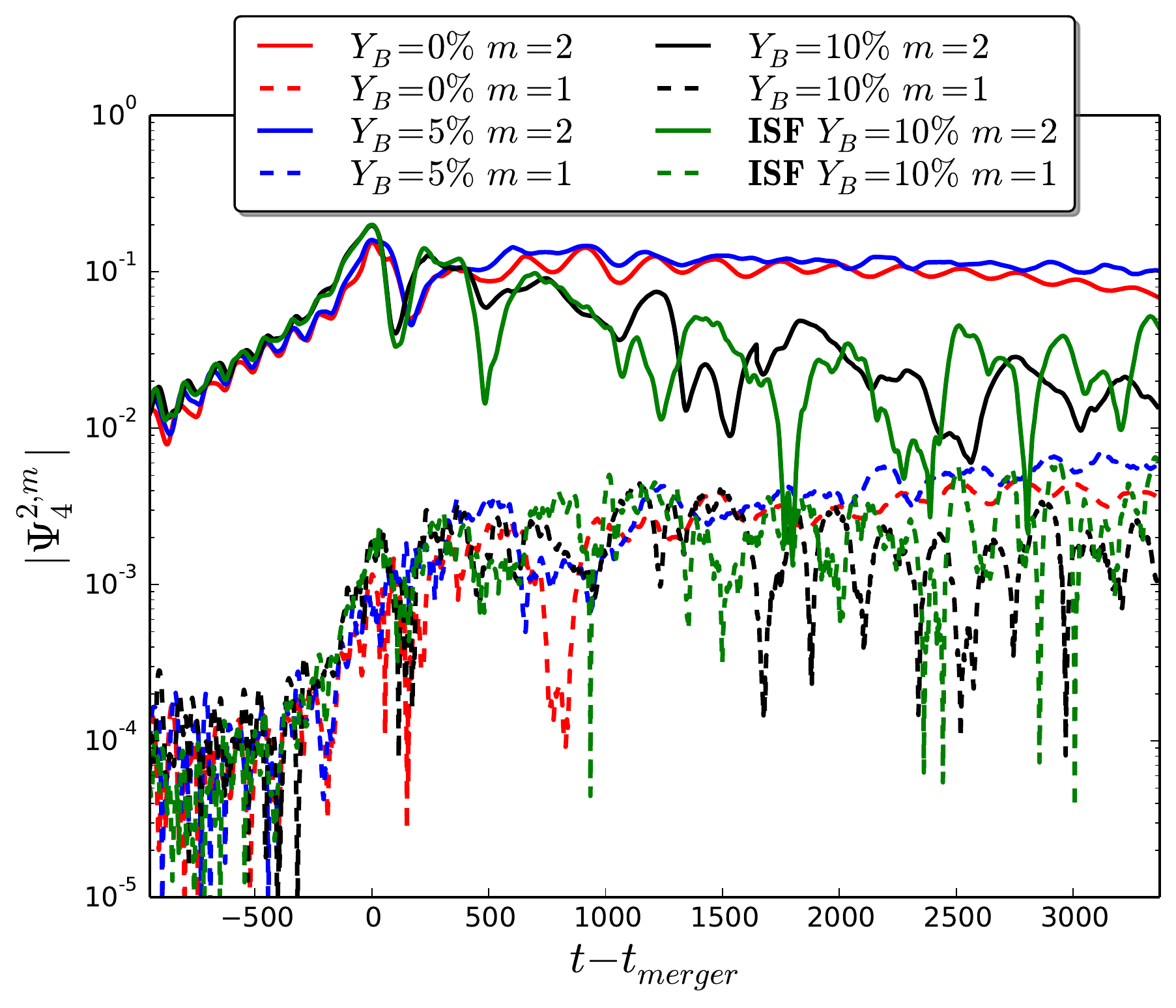}
\caption{{\em Gravitational waves}. The norm of the modes $(l,|m|)=(2,1)$ and $(l,|m|)=(2,2)$ of $|\Psi_{4}|$ as a function of time from the merger for the different cases.}
\label{norm_psi_fb}
\end{figure}

To analyze in detail this symmetry breaking, the amplitude of the $|m|=1$ and $|m|=2$ modes (both for $l=2$) of the scalar $\Psi_{4}$ are displayed in Fig.~\ref{norm_psi_fb} for all cases. As it can be observed, the amplitude of the mode $|m|=1$ has a similar behavior for all the cases, achieving roughly the same saturation level after the merger.  However, significant differences arise in the mode $|m|=2$: while the strength of this mode is roughly constant for the cases $Y_{B}=\{0\%,5\%\}$, the cases with $Y_{B}=10\%$ (both NISF and ISF) display an exponential decay soon after the merger.

\begin{figure}[h!]
\centering
\includegraphics[width=1.03\linewidth]{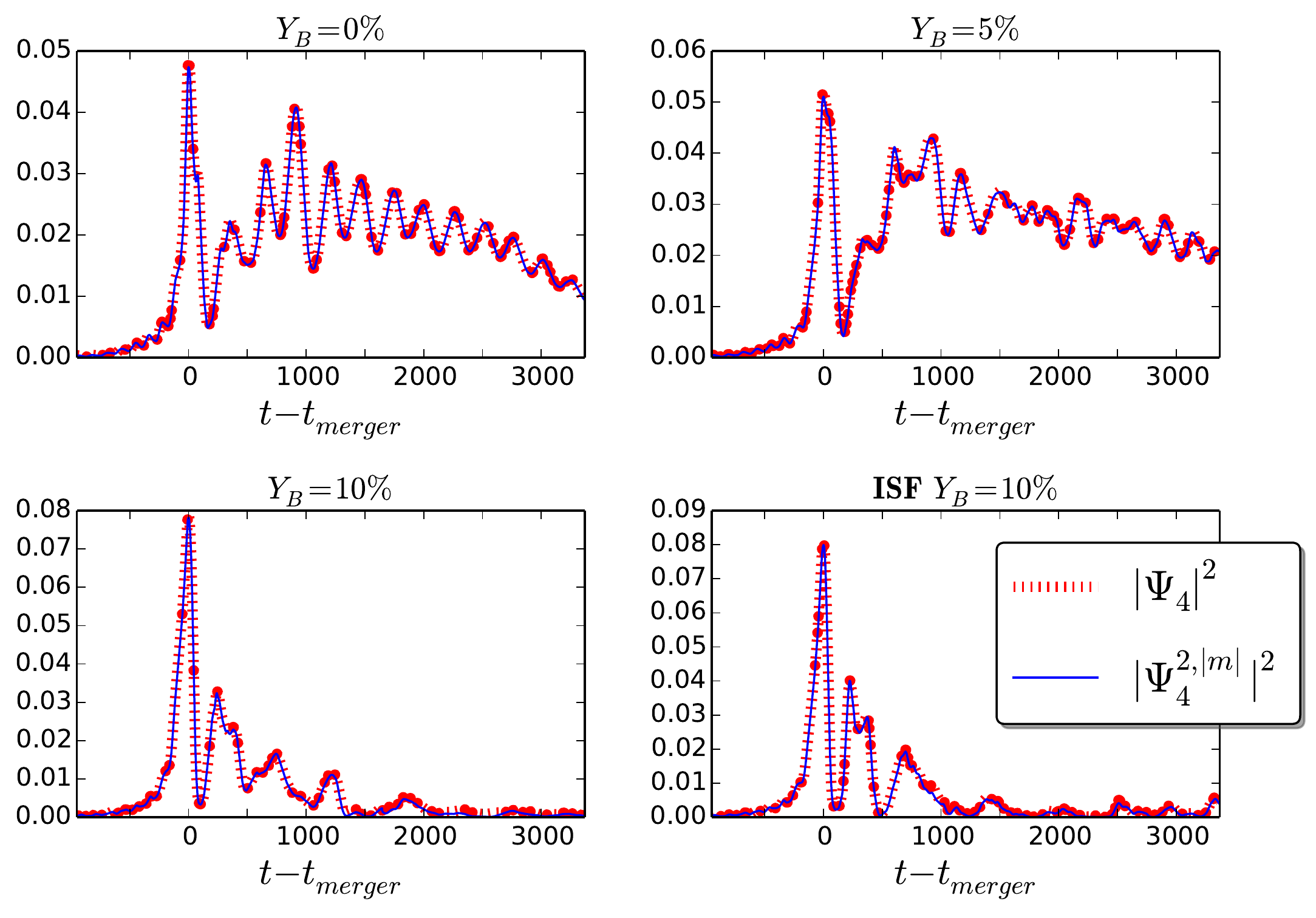}
\caption{{\em Gravitational waves.}  The norm of the total gravitational radiation emitted during the coalescence and the norm of the $l=|m|=2$ dominant modes, as a function of time from merger, for the different cases. They basically overlap, showing that the main contribution to GW emission always comes from the $l=|m|=2$ mode.}
\label{psi_all}
\end{figure}

In order to check if there is any transfer of energy through others modes, the strength of the total gravitational radiation is compared with the main mode $l=|m|=2$ in Fig.~\ref{psi_all}. Clearly, the predominant radiative contribution comes from always the main mode. Thus, we can conclude that the loss of quadrupole symmetry does not induce a significant additional radiation in other modes with other $l$ and $m$.

\begin{figure}[h!]
\centering
\includegraphics[width=1.0\linewidth]{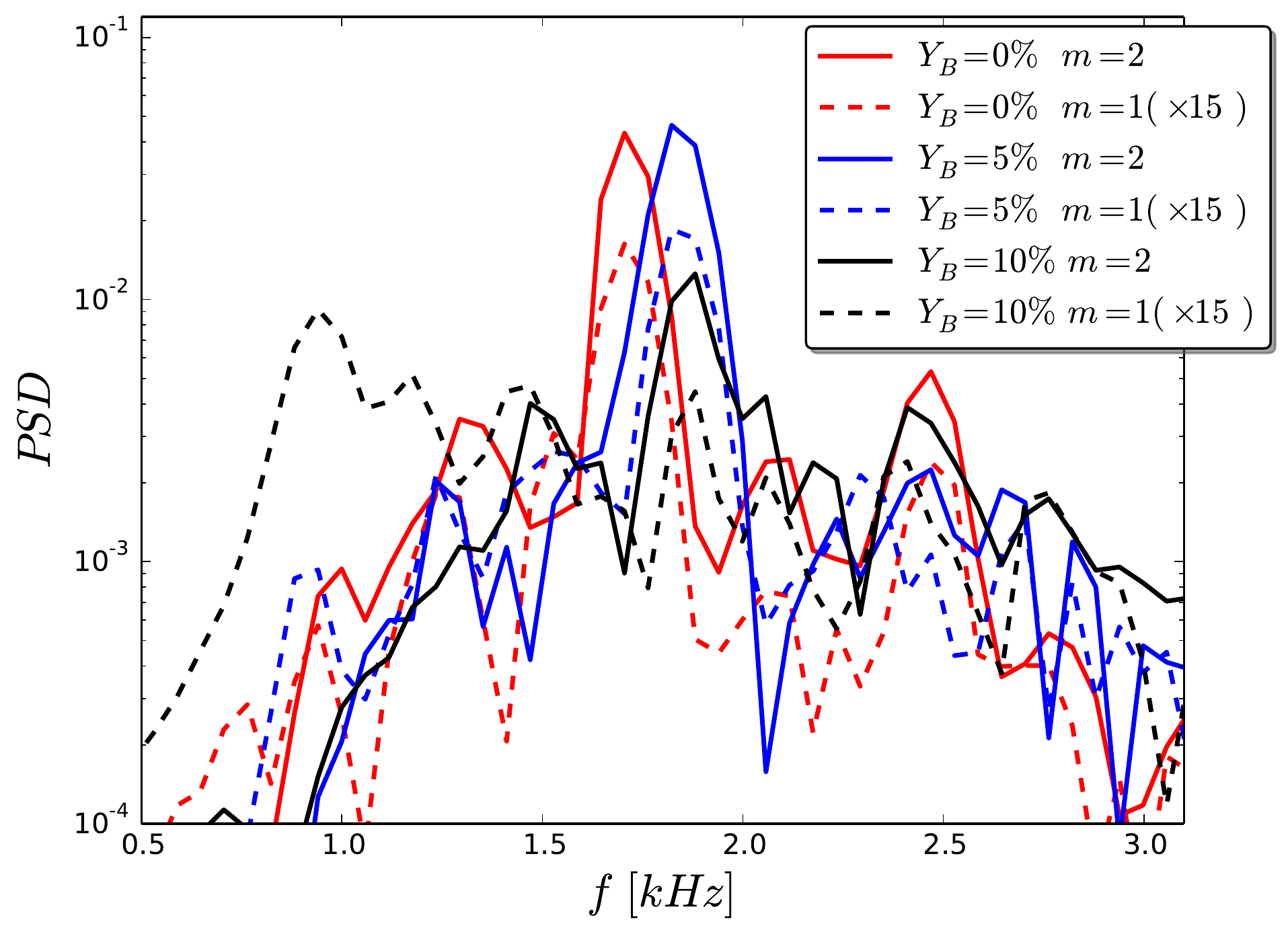}
\caption{{\em Gravitational waves.} Fourier transform of the real part of $\Psi_{4}$, considering from the merger time on, for the models NS and NISF with $Y_{B}=\{5\%,10\%\}$. We show the $(l=2,|m|=1)$ mode, amplified by a factor 15 for visualization purposes, and the dominant $(l=2,|m|=2)$ mode. The dominant radiation mode after the merger is given by $(l=2,|m|=2)$, achieving peaks at frequencies $f_{\rm peak}=\{1.62,1.81,1.87\}$ kHz respectively. The only significant $(l=2,|m|=1)$ mode corresponds to the case $Y_{B}=10\%$, with a peak at $f_{\rm m=1}=0.935$ kHz, at half the frequency of the $|m|=2$ one.}
\label{psd}
\end{figure}

Finally, we can learn further information about the properties of the remnant by analyzing the power spectral density of the $\Psi^{2,|m|}_{4}$ modes, calculated integrating from the merger time on. The $|m|=2$ mode is displayed in Fig.~\ref{psd}, together with the $|m|=1$ one, amplified in amplitude by a factor fifteen for clarity. The frequency of the dominant mode corresponds to the double of the orbital period at the merger time, and it is associated to a mixture of the rotational motion and the quadrupolar structure~\cite{PhysRevD.92.044045,PhysRevD.91.124056}, with a weak dependence on the ratio $Y_B$~\cite{suspalen}. The values of these peak frequencies are presented in Table~\ref{fb_conf}.

The peak in frequency spectrum for the $|m|=1$ mode is more than one order of magnitude weaker than the $|m|=2$ mode. However, as already previously noted for unequal binary NS mergers~\cite{2016CQGra..33r4002L,m=1}, for the model with $Y_{B}=10\%$, in which the one-arm instability fully develops, the peak of the $|m|=1$ mode is at half the frequency  of the one corresponding to the $|m|=2$ mode. Quantitatively, for that case we obtain $f_{m=1}=0.935$~kHz and $f_{m=2}=1.87$~kHz, when the one-arm instability fully develops. The exact value depends on the chosen EoS, and for a realistic one, the typical value of $f_{m=2}$ are found to be between $\sim [2-3.5]$ kHz, see Refs. \cite{PhysRevD.92.044045,2016CQGra..33r4002L}. Regardless on the specific value, for frequencies corresponding to the $|m|=1$ mode, the Earth-based detectors are more sensitive and might be able to distinguish these features, for close enough events. In particular, finding a peak with $f\sim 1$ kHz in equal-mass low-spinning binaries could be a signature of DM presence.

\section{Discussion}\label{discu_fb}
In this work, we have studied by using full 3D numerical simulations, the dynamics and gravitational radiation emitted during the coalescence of binary NSs with DM clustered in their interior. These objects have been modeled by using FBSs, i.e, compact stellar objects made with a mixture of a perfect fluid and a complex scalar field. In our model, we have considered that in each star the fermionic matter interacts with the bosonic matter cores only through gravity. In particular, we have considered binaries formed for stars with the same individual ADM mass $M=1.35M_{\odot}$ and compactness $C=0.12$, but with three different boson-to-fermion ratio $Y_{B}=\{0\%,5\%,10\%\}$, to study the dependence on the amount of DM in the stars.

We have found that, during the late inspiral, both the dynamics and the GWs radiated in these three cases are roughly the same, making it very difficult to distinguish differences with respect to a canonical binary NS with $Y_{B}=0\%$. At the merger stage differences arise in the dynamics for the cases $Y_{B}=5\%$ and $Y_{B}=10\%$: while the NSs merge and form a rotating remnant, the boson components keep orbiting maintaining its individual shape for longer times. In the late post-merger differences grow larger with respect to the case $Y_{B}=0\%,$ where the dynamics is governed by a massive NS which rotates differentially with a dominant quadrupolar shape, and producing GWs in the $l=|m|=2$ modes. In the case $Y_{B}=10\%,$ the dark bosonic cores cause a redistribution of the fermionic matter, breaking the quadrupolar symmetry of the remnant and forming an $|m|=1$ over-density through the one-arm instability, which is excited by the asymmetries introduced by the three-body interaction (i.e., fermionic matter plus two bosonic cores coupled only through gravity). In this case, the dominant GW mode $l=|m|=2$ decay exponentially much faster than for $Y_{B}=0\%$. For comparison purposes, we have also considered a binary where the bosonic DM interacts through both gravity and scalar field interactions, obtaining roughly the same results as before.  This seems to indicate that the one-arm spiral instability develops generically in the merger of NSs with DM cores due to the many-body-interaction after the merger. Note that this instability is able to break the even-mode symmetry in the density distributions, appearing qualitatively similar to the standing accretion shock instability observed in 3D supernova simulations \cite{burrows06,iwakami09}.

Let us stress the differences of our results with respect to the ones obtained in Ref.~\cite{ELLIS2018607}, where they introduce a Lagrangian with four coupled objects (i.e., two NSs and two DM cores) to describe the post-merger dynamics. They pointed out the presence of supplementary peaks at higher frequencies than the $|m|=2$ mode in the post-merger spectrum of NSs mergers, but could not anticipate the lower frequency peak due to the one-arm instability. 

As it was also notice in~\cite{m=1}, although the $|m|=1$ modes strength in our case is at least fifteen times smaller, it becomes more relevant as a contributor to the post-merger GW signal since it occurs at a frequency half of the dominant mode $|m|=2$, where the GW detectors are more sensitive. Notice however that there is some degeneracy, since this instability has also been observed to happen in several binary NS merger simulations~\cite{PhysRevD.94.064011}, especially
with spin and/or eccentricity~\cite{PhysRevD.93.024011,PhysRevD.92.121502} and for unequal mass  stars~\cite{2016CQGra..33r4002L,m=1}. There are two distinct features of our case with respect to those ones. First, the one-arm instability strongly develops even for equal mass no-spinning objects. Therefore, the waveform before the merger might contain enough information (i.e., the masses of the stars) to break partially the degeneracy and discard some of the asymmetries which could produce a strong one-arm instability.
Second, although the exponential decay affecting the $|m|=2$ mode also occurs in unequal mass or highly spinning binaries, it shows a faster attenuation in our cases. Therefore, possible detection of these modes with current or future detectors, combined with a detailed analysis of the signal during the inspiral, could constraint the presence of DM cores inside NSs and enhance our understanding of its nature.\\

\subsection*{Acknowledgments} 
We acknowledge support from the Spanish Ministry of Economy and Competitiveness grants FPA2013-41042-P and AYA2016-80289-P (AEI/FEDER, UE). CP also acknowledges support from the Spanish Ministry of Education and Science through a Ramon y Cajal grant. MB would like to thank CONICYT Becas Chile (Concurso Becas de Doctorado en el Extranjero) for financial support.
We thankfully acknowledge the computer resources at MareNostrum and the technical support provided by Barcelona Supercomputing Center, with the time granted through the $17^{th}$ PRACE regular call (project Tier-0 GEEFBNSM, P.I. CP) and the RES calls AECT-2018-1-0005 (P.I. DV), AECT-2019-1-0007 (P.I. DV).

\bibliographystyle{utphys}
\bibliography{biblio}

\end{document}